\documentclass[aps,twocolumn,superscriptaddress,amsmath,amssymb,longbibliography]{revtex4-1}

\usepackage{graphicx}
\usepackage{dcolumn}
\usepackage{multirow}
\usepackage{bm}
\usepackage{color}
\usepackage{txfonts}
\usepackage{hyperref}
\usepackage{soul}
\usepackage{CJK}
\usepackage{url}
\usepackage{graphicx}
\usepackage{dcolumn}
\usepackage{bm}
\newcolumntype{d}[1]{D{.}{.}{#1}}

\begin{document}
\begin{CJK*}{UTF8}{mj}

\title{Relational flexibility of network elements based on inconsistent community detection}

\author{Heetae Kim (김희태)}
\affiliation{Department of Industrial Engineering, Universidad de Talca, Curic{\'o} 3341717, Chile}%
\affiliation{Asia Pacific Center for Theoretical Physics, Pohang 37673, Korea}%

\author{Sang Hoon Lee (이상훈)}
\email[Corresponding author: ]{lshlj82@gntech.ac.kr}
\affiliation{Department of Liberal Arts, Gyeongnam National University of Science and Technology, Jinju 52725, Korea}%

\date{\today}

\begin{abstract}
Community identification of network components enables us to understand the mesoscale clustering structure of networks. A number of algorithms have been developed to determine the most likely community structures in networks. Such a probabilistic or stochastic nature of this problem can naturally involve the ambiguity in resultant community structures.
More specifically, stochastic algorithms can result in different community structures for each realization in principle.
In this study, instead of trying to ``solve'' this community degeneracy problem, we turn the tables by taking the degeneracy as a chance to quantify how strong companionship each node has with other nodes.
For that purpose, we define the concept of companionship inconsistency that indicates how inconsistently a node is identified as a member of a community regarding the other nodes. 
Analyzing model and real networks, we show that companionship inconsistency discloses unique characteristics of nodes, thus we suggest it as a new type of node centrality. 
In social networks, for example, companionship inconsistency can classify outsider nodes without firm community membership and promiscuous nodes with multiple connections to several communities. 
In infrastructure networks such as power grids, it can diagnose how the connection structure is evenly balanced in terms of power transmission.
Companionship inconsistency, therefore, abstracts individual nodes' intrinsic property on its relationship to a higher-order organization of the network.

\end{abstract}

\maketitle
\end{CJK*}


\section{\label{sec:level1}Introduction}
Community structures of networks~\cite{Porter2009,Fortunato2010} are arguably the most popular concept in investigating the mesoscale connectivity between node groups of networks, in the field of network science~\cite{NewmanBook}.
Various community detection algorithms have been developed to divide a network into communities based on modularity optimization~\cite{Newman:2004jh,Newman:2004ep,Clauset:2004dz,Fortunato:2007js,Blondel:2008do}, information theory~\cite{Rosvall:2008fi}, clique percolation~\cite{Palla:2005cj}, etc. 
The main objective of community detection algorithms is to provide a principled guideline to determine each node's community membership in a network.
The algorithms work under the assumption that the nodes inside each community are statistically better connected to each other, compared with the connection to the other parts of the network, which is basically the very definition of communities in networks.

There are many ways to classify community detection algorithms, but for our purpose we classify them dichotomously as the following. 
Deterministic community detection algorithms, by definition, produce a single community structure for given control parameters. 
On the other hand, stochastic algorithms can yield different community structures at each realization in principle (and in practice, as we will show).
In general, so far, the inconsistent result of the stochastic community detection has been taken as a kind of defect of such algorithms. 
In other words, the inconsistency (sometimes dubbed as the community degeneracy problem) has been taken as the inaccuracy of stochastic detection algorithms~\cite{Kwak:2011fb,Lancichinetti:2012kx}.

In this paper, however, we would like to argue that there is nothing wrong with the ``inconsistent'' results such stochastic algorithms produce, as it is fundamentally impossible to define the exact boundary of one's community identity in the first place.
For example, people are naturally involved in various groups of other people and the degrees of participation between those groups are different.
Such different types of participation of a node in groups can indicate crucial information on the node's social existence or influence.
Throughout our study, therefore, we directly confront the inconsistent community detection results of a stochastic algorithm and harness them instead of evading them. 
Based on ensembles of community detection results, we examine how frequently the nodes are identified as the different (or same) community members.
Applying the method to real networks in addition to model networks with prescribed communities, we show that the companionship inconsistency represents the sense of belongingness of nodes in networks and thus conveys their unique properties, in comparison to conventional centrality measures.

The paper is organized as the following. First, we introduce the concept of companionship inconsistency (CoI) and methodology in Sec.~\ref{sec:methods} with an illustrative example network.
We apply the method to model networks to investigate the characteristics of companionship inconsistency in Sec.~\ref{sec:random}.
In Sec.~\ref{sec:real}, we apply companionship inconsistency measure to various real networks and identify the roles of nodes. We summarize the results and conclude the paper with open questions and discussions in Sec.~\ref{sec:discussion}.

\section{\label{sec:methods}Methods}

\subsection{\label{subsec:measure}Companionship inconsistency}

\begin{figure*}[t]
\includegraphics[width=0.6\textwidth]{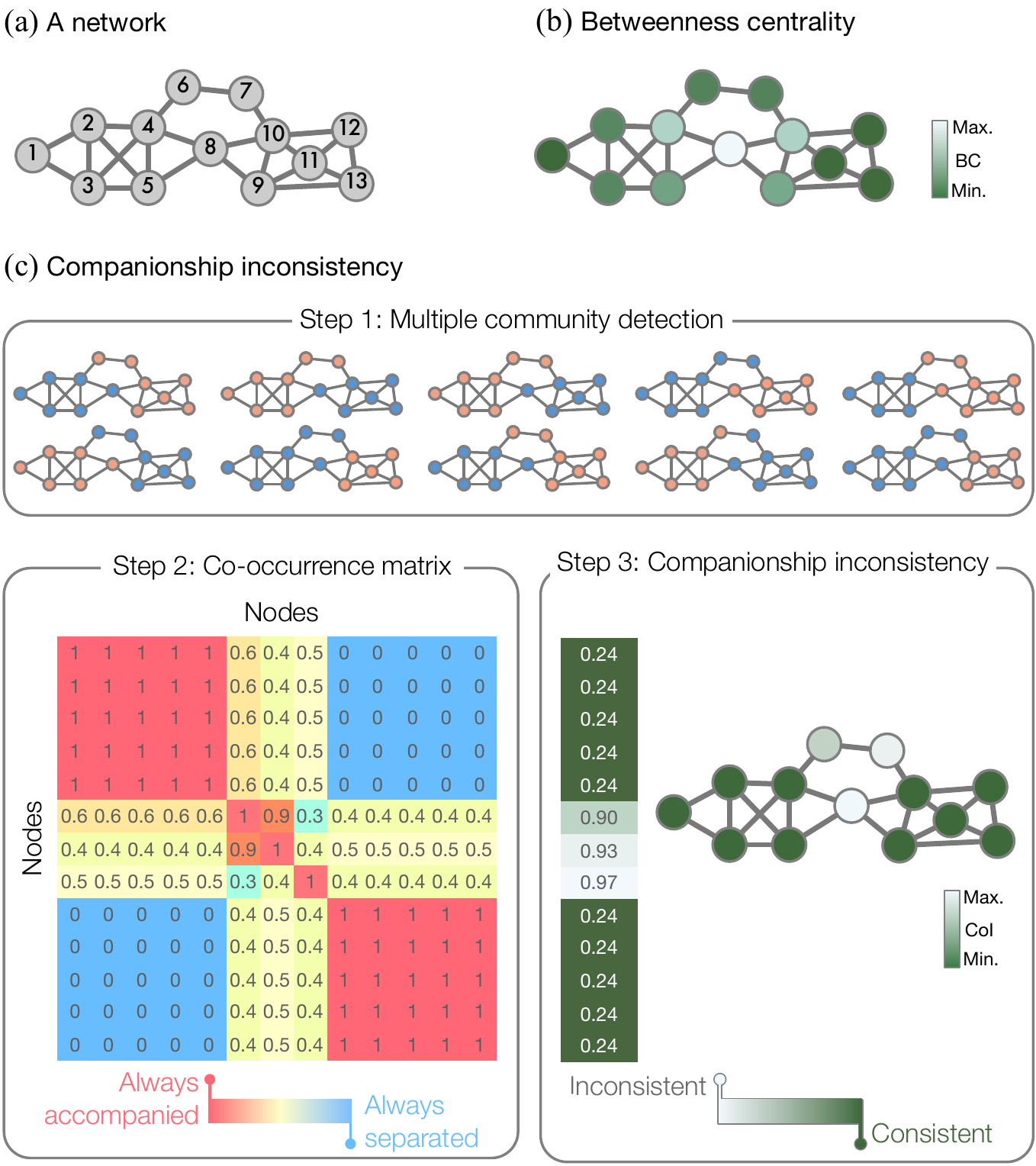}
\caption{An illustrative guide to the CoI, in particular, compared to the betweenness centrality (BC). (a) An example network. (b) The BC only counts the shortest path length likely going through the central node 8, so it cannot discern the two bridge nodes (6 and 7) in terms of community structure. (c) A schematic diagram of measuring CoI. From an ensemble of several community detection results (step 1), we extract the inconsistent community partnership (step 2). Based on this co-occurence between the nodes, we assign each node's CoI value by evaluating the overall inconsistency with the other nodes (step 3).}
\label{fig1}
\end{figure*}

In principle, the community detection algorithm based on stochastic methods may produce different results by definition, even for the results from the same parameters, and it is not difficult to observe such cases in practice. Again, we would like to emphasize that it is not only from the limitation of algorithm but also from networks' innately ambiguous characteristics when it comes to community boundaries.
To quantify the ambiguity of community structures, we introduce a principled measure of CoI for each node as a new type of centrality. The CoI captures how inconsistently the node is classified as a community's member, with or without fixed companion nodes.
When a node tends to be clustered with different nodes for different realizations, we consider that the node's community identity is inconsistent. 

In our previous paper~\cite{Kim:2015kg}, we have introduced the concept of CoI to relate the stability of power-grid nodes to their community membership structure (we defined the ``community consistency'' there, but we have changed it in this paper to focus on the inconsistent nodes representing functional flexibility). For visual illustration, see Fig.~\ref{fig1}, where we take a small example network in Fig.~\ref{fig1}(a). We recap the formal definition in this paper again for self-containedness.
To formulate the CoI, we first define the co-occurrence matrix elements
$\phi_{ij}$ as the proportion of the number of cases that nodes $i$ and $j$ are identified as the members of the same community, which corresponds to the matrix in step 2 of Fig.~\ref{fig1}(c):
\begin{equation}
\phi_{ij}=\frac{1}{n_d} \sum_{\alpha=1}^{n_d} \delta_{\alpha}(g_i,g_j) \,,
\end{equation}
where $\delta_{\alpha}$ is the Kronecker delta for the $\alpha$th realization of community detection, $g_i$ is the community index of node $i$, and $n_d$ is the total number of realizations of community detection. In other words, $\delta_{\alpha}(g_i,g_j) = 1$ when $i$ and $j$ are in the same community and $\delta_{\alpha}(g_i,g_j) = 0$ otherwise in the $\alpha$th realization of community detection. The measure $\phi_{ij}$ 
is $0$ (or $1$) if $i$ and $j$ consistently belong to the different (same) community and intermediate values when the pairwise community membership is inconsistent, respectively. The extreme values such as $\phi_{ij} = 1$ and $\phi_{ij} = 0$ represent consistency in community detection, while intermediate values represent inconsistency. Therefore, based on this, we define the CoI of node $i$ shown in step 3 of Fig.~\ref{fig1}(c), denoted by $\Phi_i$, as
\begin{equation}
\Phi_i=1-\frac{1}{N-1}\sum_{j (\neq i)}(1-2\phi_{ij})^2 \,,
\label{eq:CC}
\end{equation}
where $N$ is the total number of the nodes. As a result, $\Phi_i=0$ when node $i$ always forms communities with the same nodes. In principle, $\Phi_i=1$ implies that the probability that node $i$ is clustered with any other node is $1/2$ (the maximum uncertainty in the comembership). The ``community consistency'' defined in Ref.~\cite{Kim:2015kg} is equal to $1 - \Phi_i$. 

Note that the maximum value $\Phi_i=1$, i.e., the comembership matrix element $\phi_{ij} = 1/2$ for all of the other nodes $j$, assumes exactly two ``ground-truth'' communities potentially connected to the node $i$. Therefore, in principle, one has to be careful when it comes to the comparison of the results CoI produces, as the maximum value $\Phi_i$ can be different for specific circumstances depending on the number of communities and community size heterogeneity. For instance, the fact that $\phi_{ij}$ is a decreasing function of the number of communities attached to node $i$ can make the head-to-head comparison between the nodes attached to different numbers of communities nontrivial. In reality, however, we are not able to know the number of ground-truth communities \emph{a priori}, let alone the local communities connected to each node. Therefore, to design a measure first addressing this particular characteristic of each node, we stick to the simple assumption of two (or at least not many) communities attached to the bridge nodes. As we demonstrate in the following sections, our CoI measure produces meaningful results and works fine in practice. 

In order to measure CoI, one can utilize any stochastic community detection algorithm. 
In this study, we take the GenLouvain~\cite{Genlouvain} algorithm, which is a variant of the original Louvain algorithm~\cite{Blondel:2008do}, with the default randomization option \texttt{move}. 
The GenLouvain (just as its ancestor Louvain) algorithm separates communities based on the modularity maximization~\cite{Newman:2004jh,Newman:2004ep}.
The algorithm can detect communities in different scales by tuning the resolution parameter $\gamma$ in the modularity function~\cite{Newman:2004jh,Newman:2004ep}
\begin{equation}
Q = \frac{1}{2m} \sum_{i \ne j} \left[ \left( A_{ij} - \gamma \frac{k_i k_j}{2m} \right) \delta(g_i,g_j) \right ] \,,
\label{eq:modularity}
\end{equation}
where $A_{ij}$ is the adjacency matrix elements representing the network structure, $k_i$ is the degree (the number of neighbors) of node $i$, $g_i$ is the community index of node $i$, $\delta$ is the Kronecker delta, and $m$ is the total number of edges that plays the role of normalization factor for matching the scale of $A_{ij}$ and $k_i k_j$ terms for $\gamma = 1$, and ensuring $-1 \le Q \le 1$.
The smaller $\gamma$ values we use, the larger communities (thus the smaller number of communities in total) we detect.
To generate statistical ensembles, we run multiple realizations of the GenLouvain algorithm for given $\gamma$ values.
It general, one needs to tune the value of $\gamma$. 
For the rest of the paper, we use the $\gamma$ value in a rather heuristic way, so that it generates a reasonable number of communities, as our goal is to demonstrate the utility of CoI for various types of networks rather than provide the most precise fine-tuned value of $\gamma$ of the GenLouvain algorithm for each network.
Therefore, one always has to keep in mind that CoI values depend on the choice of different resolution parameter $\gamma$ as well. 

We note that in community detection literature, other types of measures: flexibility, promiscuity, disjointedness, cohesion strength, and Rand index also consider the change of community identity of nodes~\cite{Garcia:gz,Rand:1971ki}.
Flexibility counts the number of changing community identity of a node while promiscuity counts the number of communities to which the node ever belongs. In contrast to CoI, both flexibility and promiscuity do not consider the pairwise relationship with companions. Disjointedness and cohesion strength take the community identity of the other nodes, but disjointedness focuses on how a node independently changes its community identity apart from the other nodes and cohesion strength only counts the mutual companionship without taking the absence of companionship into count. 
The Rand index~\cite{Rand:1971ki} measures the similarity in data clusterings but it is a cluster-centric measure~\cite{Gates:2019hi}, while CoI is node centric.
The aforementioned measures also utilize the fuzziness of community~\cite{Reichardt:2004ea} as CoI does. However, those measures require the information on the community label of each node, whereas CoI only considers whether a pair of nodes are in the same community or not. Therefore, we emphasize that CoI can reveal the unique characteristics of nodes that are not captured by the seemingly similar measures, as we begin to demonstrate from now on.

\subsection{\label{subsec:concept}Implication of companionship inconsistency}

Let us revisit the example network shown in Fig.~\ref{fig1} to inspect the implication of CoI, in particular, compared to another network centrality also known to able to detect ``bridge'' nodes between groups. The network in Fig.~\ref{fig1}(a) has three nodes (denoted by 6, 7, and 8) located between two groups of nodes: the group of nodes 1, 2, 3, 4, and 5 on the left, and the other group of nodes 9, 10, 11, 12, and 13 on the right.
The nodes in the two (left and right) groups are densely connected so that they are almost always consistently detected with each group's members.
However, since the community identity of the three nodes (6, 7, and 8) is rather ambiguous, the community identity of the nodes are changed for each detection, e.g., node 6 is sometimes clustered with the left group and sometimes with the right group. Counting this co-occurrence of each node with others in detected communities, we calculate how flexible partnership the nodes have, as we presented in Sec.~\ref{subsec:measure}.

In this manner, the CoI reveals the characteristics of nodes considering the mesoscopic functional relationship between them on top of the structure. 
For instance, all three nodes 6, 7, and 8 are in the middle of the other two groups, which is captured by their large CoI values [step 3 of Fig.~\ref{fig1}(c)].
However, the portion of betweenness centrality (BC)---the fraction of shortest paths between all of the pairs of nodes that go through the node~\cite{Goh:2001ht}---is highly concentrated on node 8 and the BC values of nodes 6 and 7 are almost indistinguishable from the internal nodes such as nodes 2 and 3, because BC only takes the shortest path (likely using the path through node 8 instead of nodes 6 and 7) into account for given source and target nodes [Fig.~\ref{fig1}(b)]. 
The similar topological position of the three nodes can only be disclosed by CoI.

\begin{figure}[t]
\includegraphics[width=0.75\columnwidth]{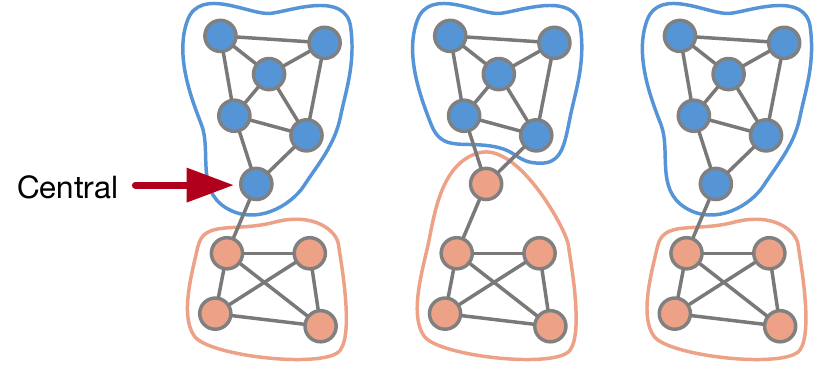}
\caption{An example showing the conceptual difference between companionship inconsistency and externality. From the three different community detection results illustrated, the central node (red arrow) has the CoI value $\Phi_{\mathrm{central}} = 8/9$ and mean externality $\langle E_{\mathrm{central}} \rangle = 4/9$.}
\label{CoIvsI}
\end{figure}

Since CoI is not solely based on the structure of the community partitioning but the mutual relationship between nodes, it reveals the functional context of nodes. 
In Fig.~\ref{CoIvsI}, for example, the central node marked by the red arrow belongs to different communities for three realizations of community detection.
The CoI of the central node is $8/9$ by calculating its comembership structure with all of the other nodes in the network.
One can see the feature of CoI clearly by comparing it to the measure called ``externality,'' which we define as
the proportion of the external degree (the number of neighbors belonging to the different community as the node of interest) of a node, i.e.,
\begin{equation}
E_i = 1-\frac{\sum_j A_{ij} \delta(g_i,g_j)}{\sum_j A_{ij}} = 1-\frac{\sum_j A_{ij} \delta(g_i,g_j)}{k_i} \,.
\label{eq:internality}
\end{equation}
It represents how many neighbors of a node belong to the different community with the node.

Compared with CoI, externality only counts the relative fraction of connections to different communities from its own. In Fig.~\ref{CoIvsI}, the central node's externality values are $1/3$, $2/3$, and $1/3$ for each realization (from the left to the right), which results in the mean externality value $\langle E_\mathrm{central} \rangle = 4/9$. By comparison, the large value of CoI focuses on the central node's functional role of switching communities, while the intermediate value of mean externality reflects the averaged level of the node's participation in other communities. In other words, the fluctuations in the membership structure is somehow ignored in the mean externality, while the CoI mainly takes such fluctuations to quantify the node's property. These two measures, as we will show later, are clearly related, but they also measure different types of bridgeness~\cite{Wu2018}.

\section{Results}
\label{sec:results}
 
\subsection{\label{sec:random}Model network}

As presented in Sec.~\ref{subsec:concept}, the CoI characterizes the attribute of nodes that cannot be captured by other simple measures without explicitly taking the community structure into account, such as BC. 
With a series of clustered network models, we find that CoI is indeed not correlated with degree (the number of each node's neighbors~\cite{NewmanBook}) or BC but externality introduced in Sec.~\ref{subsec:concept}.
We generate random subgraphs and rewire the initially internal edges into outside, to merge them into a network that consists of densely connected nodes in each group and sparsely connecting edges between the groups. 

First, we create $n_c$ number of Erd\H{o}s-R{\'e}nyi (ER) random subnetworks~\cite{ERgraph} with $N_c$ nodes and $L_c$ edges in each subnetwork with the index $c \in \{ 1, 2, \cdots, n_c \}$. There are $N=\sum_{c=1}^{n_c}{N_c}$ nodes in total, as a result. Then, we assign the initial community identity of nodes as the label of each subnetwork to which they belong and connect the nodes only to their (preassigned) community members. To set the external connections, for all of the nodes, rewire edges until $E_i=1-T_i$ (thus, $T_i = 1$ means that node $i$ is not rewired at all).
Note that as we apply the rewiring process for each node sequentially, the final value of externality of each node can be changed during the rewiring process from the other nodes. Therefore, we measure the real externality values for each network realization after all of the rewiring processes are finished. 

\begin{figure}[t]
\includegraphics[width=\columnwidth]{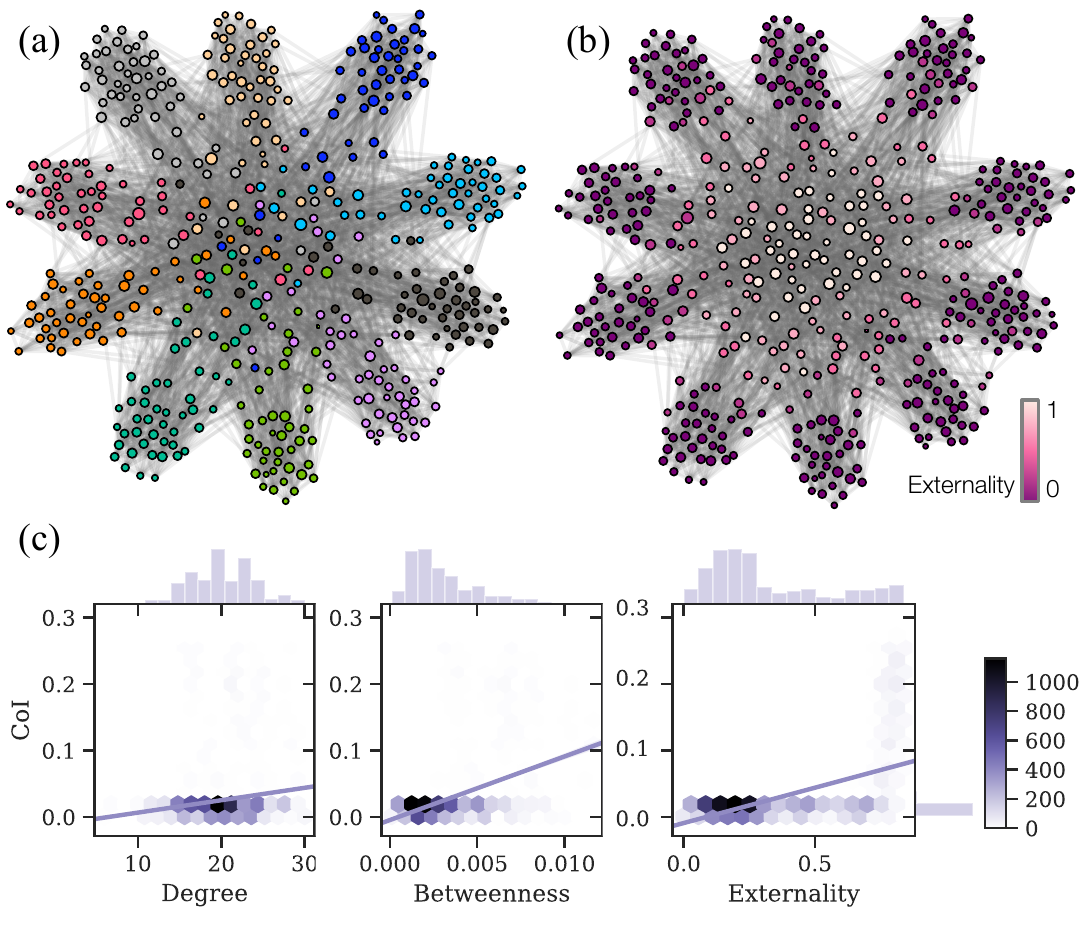}
\caption{A realization of the clustered network model with 10 communities inside, and the correlation between CoI and other network measures from $20$ network realizations and $20$ community detection results for each network. (a) The nodes belong to one of 10 communities marked by distinct colors. (b) The same network with the same layout as in panel (a) but with the different shades of color indicating the node externality value. (c) The hexagonal bin plots show the linear regression between CoI and degree, betweenness, and mean externality with the Pearson correlation coefficients $0.155$, $0.458$, and $0.553$, respectively, as reported in Table~\ref{table:correlation}. The shade of each hexagon indicates the number of corresponding data points inside each cell. The result is from $500 \times 20 = 10^4$ data points in total, as we collect all of the nodes' centrality values for $20$ different network realizations.}
\label{fig2}
\end{figure}

Figure~\ref{fig2} shows a sample network from the model and the results from an ensemble of $20$ networks composed of $n_c = 10$ subnetworks with $N_c=50$, $L_c=500$ for each of them. Therefore, the total number of nodes $N = 500$ and the total number of edges $L = 5000$ for each sample network. 
We randomly distribute the threshold value to each node independently, according to the prescribed probability distribution $p(T)$ for discrete threshold values $T \in \{ 0.2, 0.4, 0.6, 0.8, 1 \}$:
\begin{equation}
\begin{aligned}
p(T) = \frac{1}{10} & \Big[ \delta(T,0.2) + \delta(T,0.4) + \delta(T,0.6)  \\
 & + \delta(T,0.8) + 6 \delta(T,1)  \Big] \,,
\end{aligned}
\end{equation}
to generate different levels of CoI, where $\delta$ is the Kronecker delta. 
With this setting, nodes still statistically belong to their initial community (as long as $T_i > 0.1$, where $1-E_i = 1-1/n_c = 0.1$ corresponds to the case of randomly distributed membership). 
For each of this network realization, we identify $20$ community structures by independently running the GenLouvain algorithm with $\gamma = 0.7$ (that actually gives $10$ communities as we intended), and obtain CoI values of the nodes for each network.
Figure~\ref{fig2} shows a network of $20$ samples with color code based on the original cluster in Fig.~\ref{fig2}(a) and externality in Fig.~\ref{fig2}(b). 
One can identify $10$ communities of low externality nodes at the boundary and high externality nodes in the center.

\begin{table*}[t]
\caption{The Pearson correlation coefficient $r$ and $p$-value (corresponding to the null hypothesis of no correlation) between CoI and degree, betweenness, and mean externality of the clustered ER network model and real networks, where we also present the number of nodes ($N$), that of edges ($L$), and the resolution parameter $\gamma$ for community detection used. For all of the cases, the CoI and mean externality values are calculated from independent $20$ community detection results. For the clustered ER networks, the results are from the collection of the nodes' centrality values for all of the $20$ different network realizations.}
\label{table:correlation}
\begin{ruledtabular}
\begin{tabular}{cllcclll}
 Network & $N$ & $L$ & $\gamma$ & Correlation & Degree    & Betweenness & Mean externality \\ \hline
\multirow{2}{15em}{The clustered ER networks}&\multirow{2}{1em}{500}&\multirow{2}{1em}{5000}&\multirow{2}{1em}{0.7} &$r$ & $0.155$ & $0.458$   & $0.553$    \\ 
&&&&$p$-value& $9.99 \times 10^{-55}$ &$<2.22 \times 10^{-308}$   & $<2.22 \times 10^{-308}$  \\ 
\hline 
\multirow{2}{15em}{Zachary's Zachary Karate club~\cite{Zachary:1977fs}}&\multirow{2}{1em}{34}&\multirow{2}{1em}{77}&\multirow{2}{1em}{0.7}&$r$ & $-0.131$ & $-0.101$   & $0.458$   \\ 
&&&&$p$-value             & $0.461$  & $0.571$    & $6.49\times 10^{-3}$    \\\hline
\multirow{2}{15em}{Star Wars (all episodes)~\cite{StarWarssocialnet:2016iy}}&\multirow{2}{1em}{111}&\multirow{2}{1em}{450}&\multirow{2}{1em}{0.6}&$r$ & $0.311$ & $0.204$   & $0.357$   \\ 
&&&&$p$-value             & $9.06\times 10^{-4}$  & $0.0318$    & $1.22\times 10^{-4}$    \\ \hline
\multirow{2}{15em}{Work place contacts~\cite{Genois:2015gm}}&\multirow{2}{1em}{92}&\multirow{2}{1em}{755}&\multirow{2}{1em}{0.6}&$r$ & $-0.150$ & $-0.0966$   & $0.421$   \\ 
&&&&$p$-value             & $0.154$  & $0.360$    & $3.00\times 10^{-5}$    \\ \hline
\multirow{2}{15em}{Facebook friends~\cite{Mastrandrea:2015he}}&\multirow{2}{1em}{156}&\multirow{2}{1em}{4515}&\multirow{2}{1em}{0.8}&$r$ & $0.0846$ & $0.130$   & $0.542$   \\ 
&&&&$p$-value             & $0.294$  & $0.1045$    & $2.61\times 10^{-13}$    \\ \hline
\multirow{2}{15em}{Central Chilean power grid~\cite{Kim:2018ie}}&\multirow{2}{1em}{347}&\multirow{2}{1em}{444}&\multirow{2}{1em}{0.7}&$r$ & $0.0691$ & $0.0683$   & $0.332$   \\ 
&&&&$p$-value             & $0.199$  & $0.204$    & $2.4\times 10^{-10}$    \\ 
\end{tabular}
\end{ruledtabular}
\end{table*}

We compare CoI of all nodes from the $20$ sample networks with degree, betweenness, and the mean externality from $20$ realizations of GenLouvain community detection for each sample, as shown in Fig.~\ref{fig2}(c) and Table~\ref{table:correlation}.
The CoI values are weakly correlated with the degree and strongly correlated with the BC, which supposedly comes from the fact that the degree, BC, and bridgeness are all well correlated in our model networks generated from unstructured random networks. Those correlations are indeed very different in real networks as we will present in Sec.~\ref{sec:real}.

As expected, the CoI and mean externality values are (positively) well correlated. However, there is a notable difference between the two. The externality values are multimodally distributed with a wide range [as shown in the histogram on the above horizontal axis of the rightmost panel in Fig.~\ref{fig2}(c)], caused by the prescribed discrete levels of rewiring. On the other hand, most CoI values are very small and distributed within a narrow range [as shown in the histogram on the right vertical axis of the rightmost panel in Fig.~\ref{fig2}(c)], except for few outliers that are responsible for the positive correlation. 

The difference also highlights the property of CoI in comparison to externality: Even the nodes that have gone through significant amount of rewiring (e.g., the nodes with $T_i = 0.2$) still maintain their original membership profile, so their community identity itself is relatively intact, which is reflected in small CoI values even for such nodes. In contrast, the externality almost directly measures the level of rewiring, which results in the gradually changed values. In this respect, CoI is a more \emph{robust} measure to quantify the community identity of nodes with different levels of participation, compared with externality. 
In Sec.~\ref{sec:real}, we move on to real-world networks to check if these properties hold there as well.

\subsection{\label{sec:real}Real networks}

In this subsection, we examine the properties of CoI in various real networks, whose different contexts provide  opportunities to interpret CoI from multiple perspectives. For instance, if the sum of the weights attached to a node is bounded, then a larger number of neighbors (degree) of a node can result in the weaker connection assigned to each of its neighbor caused by the node's limited amount of interaction resource. In the context of social networks, the node (or a person) with a large CoI value could be an outsider, as the person's attention to its own group members will be diminished as a result. 
On the other hand, if the sum of the weights attached to a node is unbounded and scales with its degree, for instance, then a node with a large CoI value may be a multiplayer who intermediates several different communities. Therefore, practical applications may require the observation of different centralties including both CoI and conventional ones. 

We provide multiple types of such contexts, by introducing different types of networks in the following. Note that as in the case of clustered ER network model in Sec.~\ref{sec:random}, we use $20$ community detection results for each real network.

\subsubsection{\label{karate}Zachary's Zachary karate club}

\begin{figure}[t]
\includegraphics[width=\columnwidth]{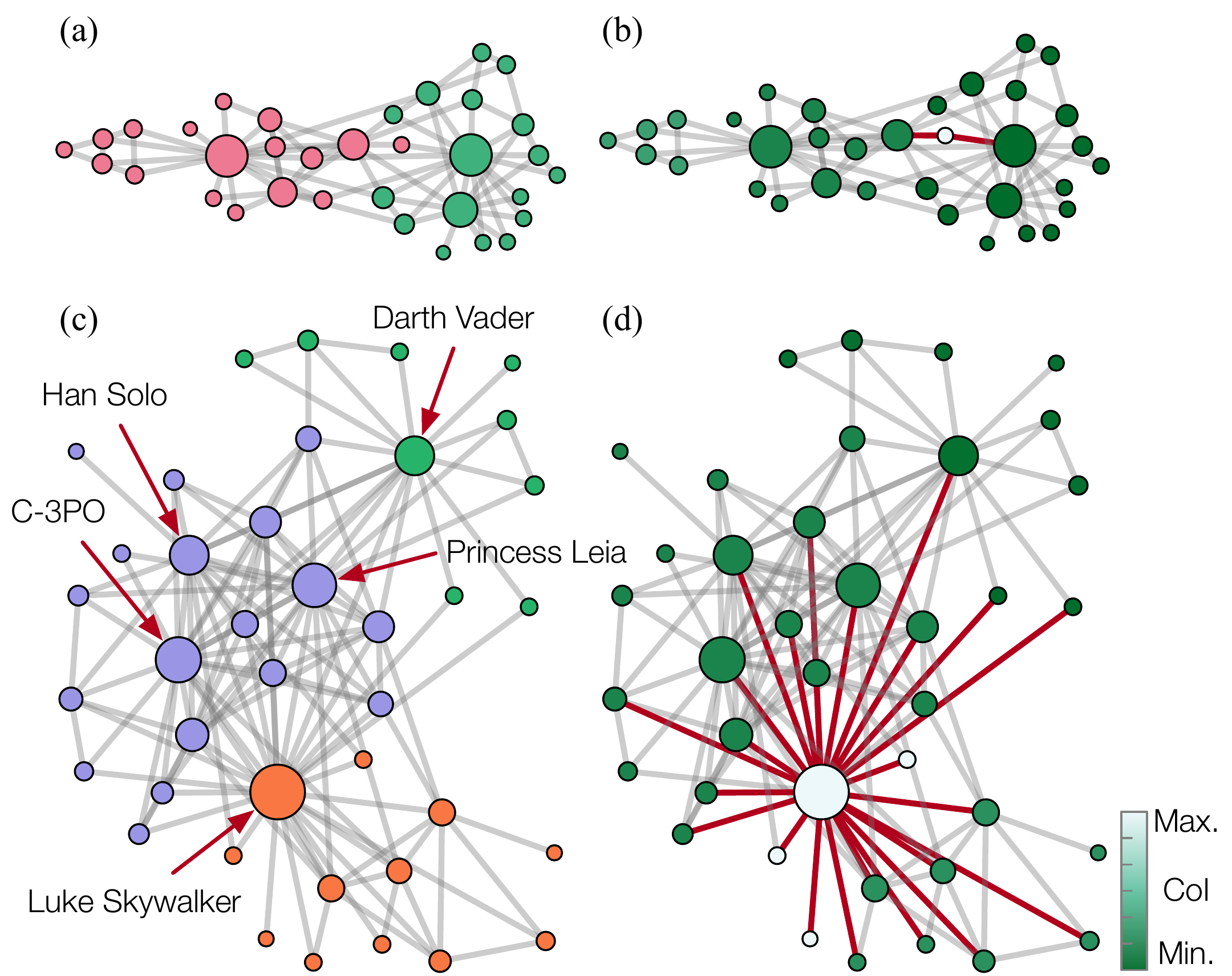}
\caption{An example of community detection result for the ZZKC network is shown in panel (a) and the CoI values from $20$ results are shown in panel (b), where the outsider node is notable with the brightest color, and the edges from the node are shown in red. Similarly, an example of community detection result for the Star Wars network is shown in panel (c) and the CoI values from $20$ results are shown in panel (d), where one of the multiplayer nodes (Luke Skywalker) is notable with the brightest color, and the edges from him are shown in red. The size of nodes is proportional to their degree, and the color shade of the nodes in panels (b) and (d) indicates the CoI values in the linearly scaled color bar at bottom right.}
\label{fig3}
\end{figure}

\begin{figure*}[t]
\includegraphics[width=\textwidth]{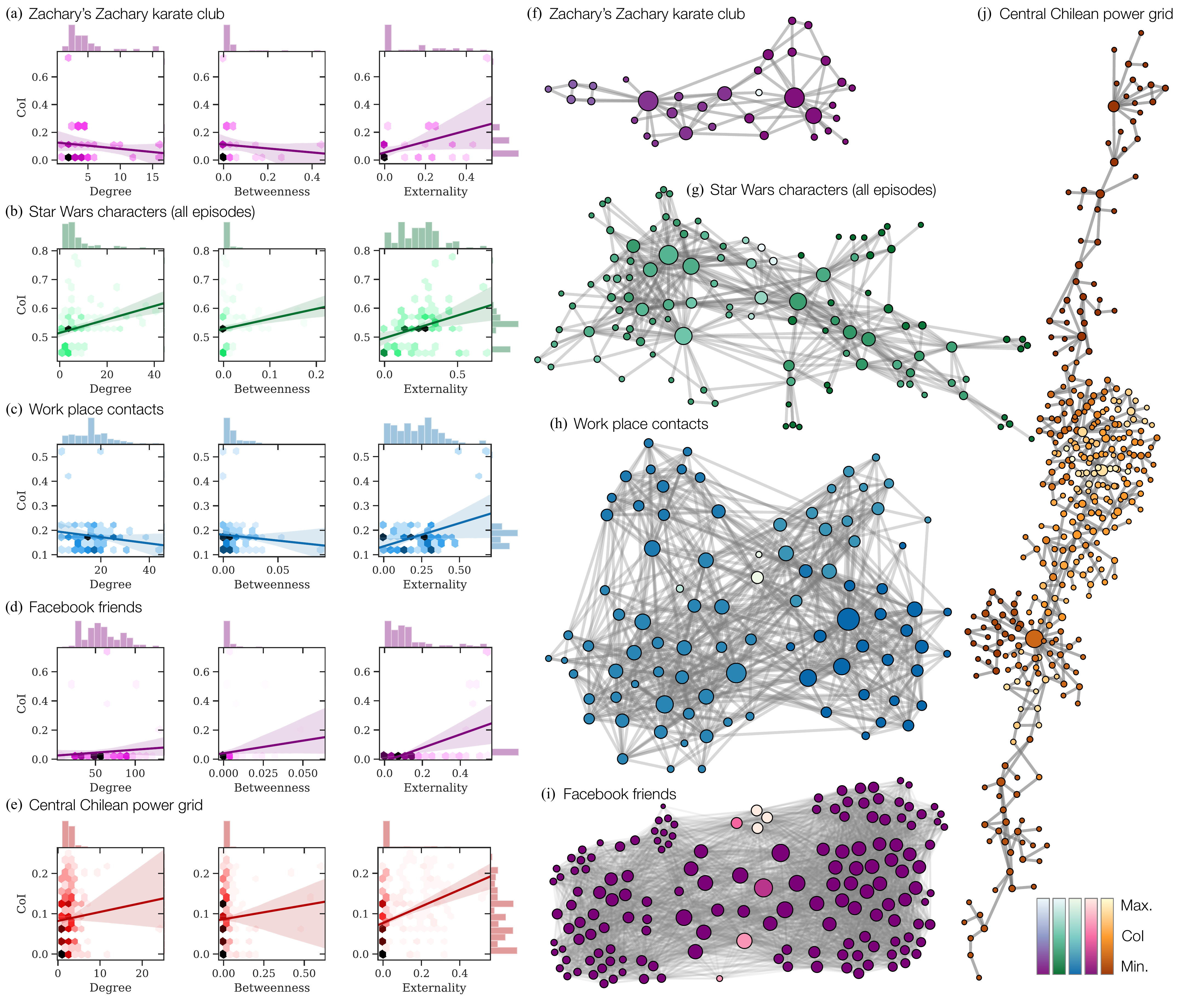}
\caption{The correlations between CoI and degree, BC, and mean externality of the real networks are presented in the left (a)--(e), and we show the networks themselves (f)--(j) with the nodes color coded by their CoI values. On the correlation plots, the hexagonal bins in each panel show the density of the data points inside. We also show the linear regression lines and the shading around them indicating the 95\% confidence interval. CoI shows the positive correlation to externality revealing the functionally distinct nodes in a network. The size of nodes is proportional to the degree. The size of nodes is proportional to their degree, and the color shade of the nodes in panels (b) and (d) indicates the linearly scaled CoI values in the color bar at bottom right.}
\label{fig4}
\end{figure*}

Zachary's karate club network~\cite{Zachary:1977fs} is the most popular benchmark network for community detection algorithms. The network is known to have two separated groups caused by the conflict between the administrator and the master. 
We use the ``original'' version of the Zachary's karate club network with two nodes with a single neighbor instead of the ``conventional'' version with one node with a single neighbor, thanks to the recent blog post by dear colleague~\cite{petterzachary}. We denote this original version by Zachary's Zachary karate club (ZZKC) network, according to the title of his blog post.

The correlations between CoI and other network centralities for the ZZKC network are listed in Table~\ref{table:correlation} and Fig.~\ref{fig4}(a) shows the regression line with the envelopes of the 95\% confidence interval on top of the scatter plot shaded by density. In contrast to the clustered ER networks in Sec.~\ref{sec:random}, the correlation between CoI and degree, and that between CoI and BC are negative, and they are not even statistically significant anyway with large $p$-values. Therefore, we can conclude that real networks with nontrivial structures would have more than just a simple positive correlation between CoI and degree or BC observed in clustered ER networks. In the ZZKC network as well, though, the correlation between CoI and mean externality is positively with statistical significance, which is not surprising considering their inherent similarity measuring community belongingness (again, the correlation is not perfect), as we argued in Sec.~\ref{subsec:concept}. 

We note that there is a peculiar type of node in this network, which we can designate as the clearly observable outsider node who is weakly connected to both sides.
As shown in Fig.~\ref{fig3}(a), the network is segregated into two communities as colored by red and green.
For multiple community detection results with $\gamma=0.7$ that yields two communities, most club members consistently belong to one of two communities.
However, there exists a person who has a notably large CoI value comparing to the others [the brightest node in Fig.~\ref{fig3}(b), which corresponds to the CoI value $>0.6$ in Fig.~\ref{fig4}(a)].
We interpret that it is caused by the fact that this outsider node is connected only to two other nodes [the red edges in Fig.~\ref{fig3}(b)], which (consistently) belong to the opposite community to each other. In particular, the node sometimes belongs to the same community with one neighbor and sometimes with the other neighbor, which results in the large value of CoI.

\subsubsection{\label{starwars}Star Wars characters}

\begin{table*}[t]
\caption{The list of Star Wars characters including a result of community partition [as illustrated in Fig.~\ref{fig3}(c)], CoI, BC, mean externality ($E$), and degree ($k$). The characters belong to one of resistances (R), Jedi knights (J), and imperial military army (M).} 
\label{StarWarsIndex}
\begin{ruledtabular}
\begin{tabular}{lclllrd{3.3}}
Character          & Community & \multicolumn{1}{c}{CoI}    & \multicolumn{1}{c}{BC} & \multicolumn{1}{c}{$E$} & \multicolumn{1}{c}{$k$} & \multicolumn{1}{c}{CoI$\times k$} \\ \hline
Luke Skywalker           &J& 0.552 & 0.3935     & 0.512       & 26     & 14.352    \\
C-3PO          &R& 0.231 & 0.1174    & 0.170        & 20     & 4.620      \\
Princess Leia           &R& 0.231 & 0.1306    & 0.274       & 19     & 4.389     \\
Han Solo            &R & 0.231 & 0.1006    & 0.144       & 16     & 3.696     \\
Darth Vader    &M& 0.189 & 0.2326    & 0.469       & 16     & 3.024     \\
R2-D2          &R& 0.231 & 0.0126    & 0.133       & 12     & 2.772     \\
Chewbacca      &R& 0.231 & 0.0117    & 0.145       & 11     & 2.541     \\
Lando          &R& 0.231 & 0.0271    & 0.227       & 11     & 2.541     \\
Wedge          &J& 0.257 & 0.0581    & 0.250        & 8      & 2.056     \\
Biggs          &J& 0.257 & 0.0164    & 0.363       & 8      & 2.056     \\
Obi-Wan        &R& 0.231 & 0.0129    & 0.287       & 8      & 1.848     \\
Mon Mothma     &R& 0.231 & 0.0005    & 0.088       & 8      & 1.848     \\
Red Leader     &J& 0.257 & 0.0136     & 0.286       & 7      & 1.799     \\
Boba Fett      &R& 0.231 & 0.0129    & 0.257       & 7      & 1.617     \\
Admiral Ackbar &R& 0.231 & 0.0065    & 0.229       & 7      & 1.617     \\
Jabba          &R& 0.231 & 0.0050    & 0.117       & 6      & 1.386     \\
Gold Leader    &J& 0.257 & 0.0009    & 0.040        & 5      & 1.285     \\
Beru           &R& 0.231 & 0.0008     & 0.140        & 5      & 1.155     \\
Yoda           &J& 0.552 & 0           & 0.350        & 2      & 1.104     \\
Zev            &J& 0.552 & 0           & 0.350        & 2      & 1.104     \\
Boushh         &R& 0.231 & 0.0006     & 0           & 4      & 0.924     \\
Owen           &R& 0.231 & 0           & 0.175       & 4      & 0.924     \\
Rieekan        &R& 0.231 & 0           & 0           & 4      & 0.924     \\
Dondonna        &J& 0.257 & 0           & 0.067       & 3      & 0.771     \\
Piett          &M& 0.189 & 0.0021    & 0.225       & 4      & 0.756     \\
Bib Fortuna    &R& 0.231 & 0           & 0.232       & 3      & 0.693     \\
Tarkin         &M& 0.189 & 0           & 0.300         & 3      & 0.567     \\
Motti          &M& 0.189 & 0           & 0.300         & 3      & 0.567     \\
Dack           &J& 0.552 & 0           & 0           & 1      & 0.552     \\
Camie          &J& 0.257 & 0           & 0.100         & 2      & 0.514     \\
Red Ten        &J& 0.257 & 0           & 0.100         & 2      & 0.514     \\
Derlin         &R& 0.231 & 0           & 0           & 2      & 0.462     \\
Ozzel          &M& 0.189 & 0           & 0         & 2      & 0.378     \\
Needa          &M& 0.189 & 0           & 0           & 2      & 0.378     \\
Emperor        &M& 0.180  & 0           & 0.500         & 2      & 0.360      \\
Anakin         &M& 0.180  & 0           & 0.500         & 2      & 0.360      \\
Janson         &J& 0.257 & 0           & 0         & 1      & 0.257     \\
Greedo         &R& 0.231 & 0           & 0         & 1      & 0.231     \\
Jerjerrod      &M& 0.189 & 0           & 0         & 1      & 0.189    
\end{tabular}
\end{ruledtabular}
\end{table*}

In social networks, CoI can also effectively identify multiplayers connecting different communities.
For instance, the Star Wars network~\cite{StarWarssocialnet:2016iy} is a social network between the characters of the movie Star Wars series. 
The nodes represent characters and the edges connect characters who communicate (not necessarily in a human language, considering R2-D2 and Chewbacca) to each other in a same scene. 
Figures~\ref{fig3}(c) and~\ref{fig3}(d) show the result of community detection and CoI of Star Wars original trilogy series for brevity (with $\gamma=0.6$ that gives three communities, roughly corresponding to the three major groups in the movie). 
The characters belong to one of the communities: resistances (violet), Jedi knights (orange), and imperial military army (green) (see Table~\ref{StarWarsIndex} for the complete list).
Luke Skywalker, C-3PO, Princess Leia, Han Solo, and Darth Vader are leading characters in the series, characterized by their large degree values. All of these leading characters interact with many other characters, but their communication is usually focused on their own communities, with an important except of Luke Skywalker.

Luke Skywalker, as the main protagonist of the original trilogy, contacts with characters from all of the three communities [the red edges in Fig.~\ref{fig3}(d)]: resistances, Jedi knights, and even from imperial military army (his father).
As a result, he achieves quite a unique status of having the largest values of both CoI and degree. As an illustrative example, we provide the list of the product of CoI and degree as a type of influence score, along with other centralities and community identity in Table~\ref{StarWarsIndex}, and Luke Skywalker's influence score is absolutely dominant compared with the others. In contrast to the outsider in ZZKC who is not necessarily influential limited by the small number of connections, we denote this type of nodes with both large CoI and degree values by ``multiplayers'' who actively connect different communities, and Luke Skywalker in this network is a representative case.
This shows that CoI, possibly combined with other network centralities, can be a useful ingredient to quantify nodes' influence.

Analyzing social relationships is not trivial~\cite{jones-correa_2012} but important, as social metrics are developed to catch bullying behaviors in school or work place~\cite{Cowie:2002bc,Alivernini:2017bl,VivoloKantor:2014cca} or to indicate influencers~\cite{Peerindex,SproutSocial}. In doing so, CoI has its merit as one requires only the information of whether a pair of people are in the same group or not, not the further information on the identities of the groups. It means that CoI needs less amount of information to analyze the companionship structure, which is particularly important considering privacy issues. We believe that CoI can augment those metrics by providing a new type of information in regard to nodes' social belonging. 

For more statistically sound results, we show the correlations between centralities for the Star Wars network of all six episodes (episodes IV, V, VI, I, II, and III, in the chronological order of the release date) with $\gamma=0.7$ in Fig.~\ref{fig4}(b).
As shown in Table~\ref{table:correlation}, again, the positive correlation between CoI and mean externality is significant, while the correlation between CoI and degree or BC is moderately positive, possibly related to Luke Skywalker's dominance for those centralities as well.

\subsubsection{\label{workplace}Work place contacts}
The work place contacts network~\cite{Genois:2015gm} represents the face-to-face contact between people in a work place, recorded by radio-frequency identification (RFID) devices. 
The place is composed of five departments in an office building, and the RFID devices tracked the contacts for 10 days. Each individual contact with the time stamp forms an edge in temporal network~\cite{TemporalNetwork}, but we aggregate all temporal edges into a static network for the purpose of our analysis.

In this network, people who play the intermediary role is not necessarily ``hub'' nodes with many neighbors, as shown in Fig.~\ref{fig4}(c) and the lack of statistically significant correlation between CoI and degree as listed in Table~\ref{table:correlation} (they show the results with $\gamma = 0.6$). Those are nodes with relatively small numbers of neighbors but connect multiple communities, which results in the large values of CoI. Compared with the ZZKC and the Star Wars network, these intermediary nodes are somewhere between the extreme outsider in ZZKC and the extreme multiplayer (Luke Skywalker) in Star Wars. We can see the positive correlation between CoI and mean externality again in this network in Table~\ref{table:correlation}.

\subsubsection{\label{facebook}Facebook friends}
The Facebook friends network~\cite{Mastrandrea:2015he} is an online social network between high school students. In contrast to the work place contacts network, the nodes in this network have more diverse range of degree values, as expected from the nature of online contacts. Except for that difference, overall, the profile of different centralities is similar to the work place contacts, with a few notable intermediary nodes [see Fig.~\ref{fig4}(d), which shows the results with $\gamma = 0.8$]. The similarity is also reflected in the lack of significant correlations between CoI and degree or BC, and the significant positive correlation between CoI and mean externality in Table~\ref{table:correlation}.

\subsubsection{\label{powergrid}Central Chilean power grid}
The central Chilean power grid is the electrical power transmission grid of the central region in Chile. 
A power grid is one of the infrastructure networks that are spatially embedded with geographical coordinates. 
The electrical power system facilities such as power plants and substations are represented as the nodes and the edges represent the high voltage transmission lines between the nodes. 
In this study, we use the ``without tap'' (WOT) version of the Chilean power-grid network~\cite{Kim:2018ie}. 
Basically, the WOT version taking it into account that the power-grid nodes are directly connected to the transmission line. By using WOT version, one can simplify the real network but it still retains the physical connection structure of the power grid (see the data description in Ref.~\cite{Kim:2018ie} for more detailed information).

The structure of power grids is determined by multiple factors such as the population, the location of natural resources, economic and environmental constraints, the distance between the facilities, etc.
On top of that, by far the most important principle is that power grids should be stably operated under possible external perturbation caused by natural disasters and intentional attacks against the infrastructure.
We have already checked the high CoI nodes are unstable against external perturbations in terms of the synchronous dynamic stability in Ref.~\cite{Kim:2015kg}, where we used a more primitive version (in terms of data processing) of the Chilean power grid than the more sophisticated version we use in this work.

By analyzing the CoI values of the WOT version in this paper, we get the additional hint about the organizational principle of power grids based on the results (with $\gamma =0.7$) shown in Fig.~\ref{fig4}(e). In contrast to the other networks used in this study with relatively narrowly distributed CoI values, the CoI values for the power grid are broadly distributed. Furthermore, if we look at the CoI values spatially distributed in the power grid (the rightmost network in Fig.~\ref{fig4}, where the node layout comes from the actual geographical coordinates), then the CoI values are gradually changed along the power-grid nodes. This gradual change indicates that the power grid is organized hierarchically in different levels, from the most local to the most global ones. The large CoI values correspond to
a few nodes in the southern region, where they connect the central and southern regions.

For the statistical correlation of CoI and other centrality measures in the central Chilean power grid, check Table~\ref{table:correlation}, where it also shares similar properties with the other networks: no statistically significant correlation between CoI and degree or BC, and the strong positive correlation between CoI and mean externality.

\section{\label{sec:discussion}Summary and outlook}

In this study, we have extracted the nodes' companionship in terms of community identity from the inconsistent community detection results.
We have measured the CoI from the individual community relationship between the pairs of nodes.
As we have demonstrated from the clustered ER network model and some real networks including ZZKC, Star Wars characters, work place contacts, Facebook friendship, and the central Chilean power grid networks, we have shown that CoI can effectively reveal the various types of nodes' social roles: outsiders, multiplayers, and building blocks of hierarchical structures.

Considering the context of networks, one can interpret CoI in various ways and apply it to acquire further information. 
For example, outsiders and multiplayers---both tend to have high companionship inconsistency---can be distinguished by degree: outsiders with small degree and multiplayers with large degree, as we have shown in the case of ZZKC and Star Wars networks. By combining several different centrality measures as such, a new classification method can be suggested, which can be  further research topics. In this sense, CoI can be a useful nodal information as a projection of network property through the mesoscale convex lens.

A common conclusion from all of the networks is that CoI and mean externality are positively correlated, which implies 
that the nodes sparsely (densely) connected to their community members are more likely to have the inconsistent (consistent) companions. One may take this as too obvious a fact, because it fits well with our intuition about the very concept of community structures. However, we believe that it is important to validate this fact by the actual data analysis as we have done in this study. An interesting future direction could be to look for exceptions to this rule.

The CoI value depends on the free parameters involved in algorithms, of course, e.g., the selection of the resolution parameter $\gamma$ for the case of using the GenLouvain algorithm. By taking this reversely, we could actually use the CoI values to determine what would be the most reasonable choice of $\gamma$. So far, people take the parameter $\gamma$ as just the factor to tune the overall community scale, but the resultant CoI values and their distribution show more than just a scale. They reveal richer structural properties such as hierarchy, as we have demonstrated in the case of the power grid. 

Beyond the simple and static network structure we use in this study, the pairwise co-occurrence can also be measured based on the evolving connection structure of temporal networks~\cite{TemporalNetwork} by considering each time series as the individual snapshot of network for community detection. In addition, in the same manner of individually counting the multiple identity of a node, CoI can be applied to hierarchical or overlapping communities~\cite{Palla:2005cj,Ahn:2010dj}. For example, one can capture a comprehensive landscape of CoI from an ensemble of community detection results generated from different resolution parameter values~\cite{Jeub:2018ik} or even from different community detection algorithms. 
The CoI is based on the mutual companionship between a pair of nodes. However, even higher-order relations such as triplet or quadruplets~\cite{Gates:2019hi,Gates:2017ta} can also be used to analyze their consistency. All of these are nice candidates for the future study, we believe.

\begin{acknowledgments}
The authors greatly thank Petter Holme for establishing the formal definition of CoI during our collaboration~\cite{Kim:2015kg}, Yong-Yeol Ahn for the fruitful discussions, and an anonymous referee for the insightful review on the formulation of the CoI measure.
This work was supported by Gyeongnam National University of Science and Technology Grant in 2018--2019.
\end{acknowledgments}

\bibliography{main_revised}

\begin{thebibliography}{37}%
\makeatletter
\providecommand \@ifxundefined [1]{%
 \@ifx{#1\undefined}
}%
\providecommand \@ifnum [1]{%
 \ifnum #1\expandafter \@firstoftwo
 \else \expandafter \@secondoftwo
 \fi
}%
\providecommand \@ifx [1]{%
 \ifx #1\expandafter \@firstoftwo
 \else \expandafter \@secondoftwo
 \fi
}%
\providecommand \natexlab [1]{#1}%
\providecommand \enquote  [1]{``#1''}%
\providecommand \bibnamefont  [1]{#1}%
\providecommand \bibfnamefont [1]{#1}%
\providecommand \citenamefont [1]{#1}%
\providecommand \href@noop [0]{\@secondoftwo}%
\providecommand \href [0]{\begingroup \@sanitize@url \@href}%
\providecommand \@href[1]{\@@startlink{#1}\@@href}%
\providecommand \@@href[1]{\endgroup#1\@@endlink}%
\providecommand \@sanitize@url [0]{\catcode `\\12\catcode `\$12\catcode
  `\&12\catcode `\#12\catcode `\^12\catcode `\_12\catcode `\%12\relax}%
\providecommand \@@startlink[1]{}%
\providecommand \@@endlink[0]{}%
\providecommand \url  [0]{\begingroup\@sanitize@url \@url }%
\providecommand \@url [1]{\endgroup\@href {#1}{\urlprefix }}%
\providecommand \urlprefix  [0]{URL }%
\providecommand \Eprint [0]{\href }%
\providecommand \doibase [0]{http://dx.doi.org/}%
\providecommand \selectlanguage [0]{\@gobble}%
\providecommand \bibinfo  [0]{\@secondoftwo}%
\providecommand \bibfield  [0]{\@secondoftwo}%
\providecommand \translation [1]{[#1]}%
\providecommand \BibitemOpen [0]{}%
\providecommand \bibitemStop [0]{}%
\providecommand \bibitemNoStop [0]{.\EOS\space}%
\providecommand \EOS [0]{\spacefactor3000\relax}%
\providecommand \BibitemShut  [1]{\csname bibitem#1\endcsname}%
\let\auto@bib@innerbib\@empty
\bibitem [{\citenamefont {Porter}\ \emph {et~al.}(2009)\citenamefont {Porter},
  \citenamefont {Onnela},\ and\ \citenamefont {Mucha}}]{Porter2009}%
  \BibitemOpen
  \bibfield  {author} {\bibinfo {author} {\bibfnamefont {M.~A.}\ \bibnamefont
  {Porter}}, \bibinfo {author} {\bibfnamefont {J.~P.}\ \bibnamefont {Onnela}},
  \ and\ \bibinfo {author} {\bibfnamefont {P.~J.}\ \bibnamefont {Mucha}},\
  }\bibfield  {title} {\enquote {\bibinfo {title} {Communities in networks},}\
  }\href@noop {} {\bibfield  {journal} {\bibinfo  {journal} {Not. Am. Math.
  Soc.}\ }\textbf {\bibinfo {volume} {56}},\ \bibinfo {pages} {1082} (\bibinfo
  {year} {2009})}\BibitemShut {NoStop}%
\bibitem [{\citenamefont {Fortunato}(2010)}]{Fortunato2010}%
  \BibitemOpen
  \bibfield  {author} {\bibinfo {author} {\bibfnamefont {S.}~\bibnamefont
  {Fortunato}},\ }\bibfield  {title} {\enquote {\bibinfo {title} {Community
  detection in graphs},}\ }\href {\doibase
  https://doi.org/10.1016/j.physrep.2009.11.002} {\bibfield  {journal}
  {\bibinfo  {journal} {Phys. Rep.}\ }\textbf {\bibinfo {volume} {486}},\
  \bibinfo {pages} {75 -- 174} (\bibinfo {year} {2010})}\BibitemShut {NoStop}%
\bibitem [{\citenamefont {Newman}(2010)}]{NewmanBook}%
  \BibitemOpen
  \bibfield  {author} {\bibinfo {author} {\bibfnamefont {M~E~J}\ \bibnamefont
  {Newman}},\ }\href@noop {} {\emph {\bibinfo {title} {Networks: An
  Introduction}}}\ (\bibinfo  {publisher} {Oxford University Press, Inc.},\
  \bibinfo {address} {New York, NY, USA},\ \bibinfo {year} {2010})\BibitemShut
  {NoStop}%
\bibitem [{\citenamefont {Newman}(2004)}]{Newman:2004jh}%
  \BibitemOpen
  \bibfield  {author} {\bibinfo {author} {\bibfnamefont {M~E~J}\ \bibnamefont
  {Newman}},\ }\bibfield  {title} {\enquote {\bibinfo {title} {{Fast algorithm
  for detecting community structure in networks}},}\ }\href@noop {} {\bibfield
  {journal} {\bibinfo  {journal} {Phys. Rev. E}\ }\textbf {\bibinfo {volume}
  {69}},\ \bibinfo {pages} {066133} (\bibinfo {year} {2004})}\BibitemShut
  {NoStop}%
\bibitem [{\citenamefont {Newman}\ and\ \citenamefont
  {Girvan}(2004)}]{Newman:2004ep}%
  \BibitemOpen
  \bibfield  {author} {\bibinfo {author} {\bibfnamefont {M~E~J}\ \bibnamefont
  {Newman}}\ and\ \bibinfo {author} {\bibfnamefont {M}~\bibnamefont {Girvan}},\
  }\bibfield  {title} {\enquote {\bibinfo {title} {{Finding and evaluating
  community structure in networks}},}\ }\href@noop {} {\bibfield  {journal}
  {\bibinfo  {journal} {Phys. Rev. E}\ }\textbf {\bibinfo {volume} {69}},\
  \bibinfo {pages} {026113} (\bibinfo {year} {2004})}\BibitemShut {NoStop}%
\bibitem [{\citenamefont {Clauset}\ \emph {et~al.}(2004)\citenamefont
  {Clauset}, \citenamefont {Newman},\ and\ \citenamefont
  {Moore}}]{Clauset:2004dz}%
  \BibitemOpen
  \bibfield  {author} {\bibinfo {author} {\bibfnamefont {A.}~\bibnamefont
  {Clauset}}, \bibinfo {author} {\bibfnamefont {M~E~J}\ \bibnamefont {Newman}},
  \ and\ \bibinfo {author} {\bibfnamefont {C.}~\bibnamefont {Moore}},\
  }\bibfield  {title} {\enquote {\bibinfo {title} {{Finding community structure
  in very large networks}},}\ }\href@noop {} {\bibfield  {journal} {\bibinfo
  {journal} {Phys. Rev. E}\ }\textbf {\bibinfo {volume} {70}},\ \bibinfo
  {pages} {066111} (\bibinfo {year} {2004})}\BibitemShut {NoStop}%
\bibitem [{\citenamefont {Fortunato}\ and\ \citenamefont
  {Barth{\'e}lemy}(2007)}]{Fortunato:2007js}%
  \BibitemOpen
  \bibfield  {author} {\bibinfo {author} {\bibfnamefont {S.}~\bibnamefont
  {Fortunato}}\ and\ \bibinfo {author} {\bibfnamefont {M.}~\bibnamefont
  {Barth{\'e}lemy}},\ }\bibfield  {title} {\enquote {\bibinfo {title}
  {{Resolution limit in community detection}},}\ }\href@noop {} {\bibfield
  {journal} {\bibinfo  {journal} {Proc. Natl. Acad. Sci. U.S.A.}\ }\textbf
  {\bibinfo {volume} {104}},\ \bibinfo {pages} {36} (\bibinfo {year}
  {2007})}\BibitemShut {NoStop}%
\bibitem [{\citenamefont {Blondel}\ \emph {et~al.}(2008)\citenamefont
  {Blondel}, \citenamefont {Guillaume}, \citenamefont {Lambiotte},\ and\
  \citenamefont {Lefebvre}}]{Blondel:2008do}%
  \BibitemOpen
  \bibfield  {author} {\bibinfo {author} {\bibfnamefont {V.~D.}\ \bibnamefont
  {Blondel}}, \bibinfo {author} {\bibfnamefont {J.-L.}\ \bibnamefont
  {Guillaume}}, \bibinfo {author} {\bibfnamefont {R.}~\bibnamefont
  {Lambiotte}}, \ and\ \bibinfo {author} {\bibfnamefont {E.}~\bibnamefont
  {Lefebvre}},\ }\bibfield  {title} {\enquote {\bibinfo {title} {{Fast
  unfolding of communities in large networks}},}\ }\href@noop {} {\bibfield
  {journal} {\bibinfo  {journal} {J. Stat. Mech.: Theory Exp.}\ }\textbf
  {\bibinfo {volume} {2008}},\ \bibinfo {pages} {P10008} (\bibinfo {year}
  {2008})}\BibitemShut {NoStop}%
\bibitem [{\citenamefont {Rosvall}\ and\ \citenamefont
  {Bergstrom}(2008)}]{Rosvall:2008fi}%
  \BibitemOpen
  \bibfield  {author} {\bibinfo {author} {\bibfnamefont {M.}~\bibnamefont
  {Rosvall}}\ and\ \bibinfo {author} {\bibfnamefont {C.~T.}\ \bibnamefont
  {Bergstrom}},\ }\bibfield  {title} {\enquote {\bibinfo {title} {{Maps of
  random walks on complex networks reveal community structure}},}\ }\href@noop
  {} {\bibfield  {journal} {\bibinfo  {journal} {Proc. Natl. Acad. Sci.
  U.S.A.}\ }\textbf {\bibinfo {volume} {105}},\ \bibinfo {pages} {1118}
  (\bibinfo {year} {2008})}\BibitemShut {NoStop}%
\bibitem [{\citenamefont {Palla}\ \emph {et~al.}(2005)\citenamefont {Palla},
  \citenamefont {Derenyi}, \citenamefont {Farkas},\ and\ \citenamefont
  {Vicsek}}]{Palla:2005cj}%
  \BibitemOpen
  \bibfield  {author} {\bibinfo {author} {\bibfnamefont {G.}~\bibnamefont
  {Palla}}, \bibinfo {author} {\bibfnamefont {I.}~\bibnamefont {Derenyi}},
  \bibinfo {author} {\bibfnamefont {I.}~\bibnamefont {Farkas}}, \ and\ \bibinfo
  {author} {\bibfnamefont {T.}~\bibnamefont {Vicsek}},\ }\bibfield  {title}
  {\enquote {\bibinfo {title} {{Uncovering the overlapping community structure
  of complex networks in nature and society}},}\ }\href@noop {} {\bibfield
  {journal} {\bibinfo  {journal} {Nature}\ }\textbf {\bibinfo {volume} {435}},\
  \bibinfo {pages} {814} (\bibinfo {year} {2005})}\BibitemShut {NoStop}%
\bibitem [{\citenamefont {Kwak}\ \emph {et~al.}(2011)\citenamefont {Kwak},
  \citenamefont {Moon}, \citenamefont {Eom}, \citenamefont {Choi},\ and\
  \citenamefont {Jeong}}]{Kwak:2011fb}%
  \BibitemOpen
  \bibfield  {author} {\bibinfo {author} {\bibfnamefont {H.}~\bibnamefont
  {Kwak}}, \bibinfo {author} {\bibfnamefont {S.}~\bibnamefont {Moon}}, \bibinfo
  {author} {\bibfnamefont {Y.-H.}\ \bibnamefont {Eom}}, \bibinfo {author}
  {\bibfnamefont {Y.}~\bibnamefont {Choi}}, \ and\ \bibinfo {author}
  {\bibfnamefont {H.}~\bibnamefont {Jeong}},\ }\bibfield  {title} {\enquote
  {\bibinfo {title} {{Consistent community identification in complex
  networks}},}\ }\href@noop {} {\bibfield  {journal} {\bibinfo  {journal} {J.
  Kor. Phys. Soc.}\ }\textbf {\bibinfo {volume} {59}},\ \bibinfo {pages} {3128}
  (\bibinfo {year} {2011})}\BibitemShut {NoStop}%
\bibitem [{\citenamefont {Lancichinetti}\ and\ \citenamefont
  {Fortunato}(2012)}]{Lancichinetti:2012kx}%
  \BibitemOpen
  \bibfield  {author} {\bibinfo {author} {\bibfnamefont {A.}~\bibnamefont
  {Lancichinetti}}\ and\ \bibinfo {author} {\bibfnamefont {S.}~\bibnamefont
  {Fortunato}},\ }\bibfield  {title} {\enquote {\bibinfo {title} {{Consensus
  clustering in complex networks}},}\ }\href@noop {} {\bibfield  {journal}
  {\bibinfo  {journal} {Sci. Rep.}\ }\textbf {\bibinfo {volume} {2}},\ \bibinfo
  {pages} {336} (\bibinfo {year} {2012})}\BibitemShut {NoStop}%
\bibitem [{\citenamefont {Kim}\ \emph {et~al.}(2015)\citenamefont {Kim},
  \citenamefont {Lee},\ and\ \citenamefont {Holme}}]{Kim:2015kg}%
  \BibitemOpen
  \bibfield  {author} {\bibinfo {author} {\bibfnamefont {H.}~\bibnamefont
  {Kim}}, \bibinfo {author} {\bibfnamefont {S.~H.}\ \bibnamefont {Lee}}, \ and\
  \bibinfo {author} {\bibfnamefont {P.}~\bibnamefont {Holme}},\ }\bibfield
  {title} {\enquote {\bibinfo {title} {{Community consistency determines the
  stability transition window of power-grid nodes}},}\ }\href@noop {}
  {\bibfield  {journal} {\bibinfo  {journal} {New J. Phys.}\ }\textbf {\bibinfo
  {volume} {17}},\ \bibinfo {pages} {113005} (\bibinfo {year}
  {2015})}\BibitemShut {NoStop}%
\bibitem [{\citenamefont {Jeub}\ \emph {et~al.}(2011-2017)\citenamefont {Jeub},
  \citenamefont {Bazzi}, \citenamefont {Jutla},\ and\ \citenamefont
  {Mucha}}]{Genlouvain}%
  \BibitemOpen
  \bibfield  {author} {\bibinfo {author} {\bibfnamefont {L.~G.~S.}\
  \bibnamefont {Jeub}}, \bibinfo {author} {\bibfnamefont {M.}~\bibnamefont
  {Bazzi}}, \bibinfo {author} {\bibfnamefont {I.~S.}\ \bibnamefont {Jutla}}, \
  and\ \bibinfo {author} {\bibfnamefont {P.~J.}\ \bibnamefont {Mucha}},\ }\href
  {https://github.com/GenLouvain/GenLouvain} {\enquote {\bibinfo {title} {{A
  generalized Louvain method for community detection implemented in MATLAB}},}\
  }\bibinfo {howpublished} {\url{https://github.com/GenLouvain/GenLouvain}}
  (\bibinfo {year} {2011-2017})\BibitemShut {NoStop}%
\bibitem [{\citenamefont {Garcia}\ \emph {et~al.}(2018)\citenamefont {Garcia},
  \citenamefont {Ashourvan}, \citenamefont {Muldoon}, \citenamefont {Vettel},\
  and\ \citenamefont {Bassett}}]{Garcia:gz}%
  \BibitemOpen
  \bibfield  {author} {\bibinfo {author} {\bibfnamefont {J.~O}\ \bibnamefont
  {Garcia}}, \bibinfo {author} {\bibfnamefont {A.}~\bibnamefont {Ashourvan}},
  \bibinfo {author} {\bibfnamefont {S.}~\bibnamefont {Muldoon}}, \bibinfo
  {author} {\bibfnamefont {J.~M.}\ \bibnamefont {Vettel}}, \ and\ \bibinfo
  {author} {\bibfnamefont {D.~S.}\ \bibnamefont {Bassett}},\ }\bibfield
  {title} {\enquote {\bibinfo {title} {{Applications of community detection
  techniques to brain graphs: Algorithmic considerations and implications for
  neural function}},}\ }\href@noop {} {\bibfield  {journal} {\bibinfo
  {journal} {Proc. IEEE}\ }\textbf {\bibinfo {volume} {106}},\ \bibinfo {pages}
  {846} (\bibinfo {year} {2018})}\BibitemShut {NoStop}%
\bibitem [{\citenamefont {Rand}(1971)}]{Rand:1971ki}%
  \BibitemOpen
  \bibfield  {author} {\bibinfo {author} {\bibfnamefont {W.~M.}\ \bibnamefont
  {Rand}},\ }\bibfield  {title} {\enquote {\bibinfo {title} {{Objective
  criteria for the evaluation of clustering methods}},}\ }\href@noop {}
  {\bibfield  {journal} {\bibinfo  {journal} {J.Am. Stat. Assoc.}\ }\textbf
  {\bibinfo {volume} {66}},\ \bibinfo {pages} {846} (\bibinfo {year}
  {1971})}\BibitemShut {NoStop}%
\bibitem [{\citenamefont {Gates}\ \emph {et~al.}(2019)\citenamefont {Gates},
  \citenamefont {Wood}, \citenamefont {Hetrick},\ and\ \citenamefont
  {Ahn}}]{Gates:2019hi}%
  \BibitemOpen
  \bibfield  {author} {\bibinfo {author} {\bibfnamefont {A.~J.}\ \bibnamefont
  {Gates}}, \bibinfo {author} {\bibfnamefont {I.~B.}\ \bibnamefont {Wood}},
  \bibinfo {author} {\bibfnamefont {W.~P.}\ \bibnamefont {Hetrick}}, \ and\
  \bibinfo {author} {\bibfnamefont {Y.-Y.}\ \bibnamefont {Ahn}},\ }\bibfield
  {title} {\enquote {\bibinfo {title} {{Element-centric clustering comparison
  unifies overlaps and hierarchy}},}\ }\href@noop {} {\bibfield  {journal}
  {\bibinfo  {journal} {Sci. Rep.}\ }\textbf {\bibinfo {volume} {9}},\ \bibinfo
  {pages} {8574} (\bibinfo {year} {2019})}\BibitemShut {NoStop}%
\bibitem [{\citenamefont {Reichardt}\ and\ \citenamefont
  {Bornholdt}(2004)}]{Reichardt:2004ea}%
  \BibitemOpen
  \bibfield  {author} {\bibinfo {author} {\bibfnamefont {J.}~\bibnamefont
  {Reichardt}}\ and\ \bibinfo {author} {\bibfnamefont {S.}~\bibnamefont
  {Bornholdt}},\ }\bibfield  {title} {\enquote {\bibinfo {title} {{Detecting
  fuzzy community structures in complex networks with a Potts model}},}\
  }\href@noop {} {\bibfield  {journal} {\bibinfo  {journal} {Phys. Rev. Lett.}\
  }\textbf {\bibinfo {volume} {93}},\ \bibinfo {pages} {218701} (\bibinfo
  {year} {2004})}\BibitemShut {NoStop}%
\bibitem [{\citenamefont {Goh}\ \emph {et~al.}(2001)\citenamefont {Goh},
  \citenamefont {Kahng},\ and\ \citenamefont {Kim}}]{Goh:2001ht}%
  \BibitemOpen
  \bibfield  {author} {\bibinfo {author} {\bibfnamefont {K~I}\ \bibnamefont
  {Goh}}, \bibinfo {author} {\bibfnamefont {B}~\bibnamefont {Kahng}}, \ and\
  \bibinfo {author} {\bibfnamefont {D}~\bibnamefont {Kim}},\ }\bibfield
  {title} {\enquote {\bibinfo {title} {{Universal behavior of load distribution
  in scale-free networks}},}\ }\href@noop {} {\bibfield  {journal} {\bibinfo
  {journal} {Phys. Rev. Lett.}\ }\textbf {\bibinfo {volume} {87}},\ \bibinfo
  {pages} {278701} (\bibinfo {year} {2001})}\BibitemShut {NoStop}%
\bibitem [{\citenamefont {Wu}\ \emph {et~al.}(2018)\citenamefont {Wu},
  \citenamefont {Tian},\ and\ \citenamefont {Liu}}]{Wu2018}%
  \BibitemOpen
  \bibfield  {author} {\bibinfo {author} {\bibfnamefont {A.-K.}\ \bibnamefont
  {Wu}}, \bibinfo {author} {\bibfnamefont {L.}~\bibnamefont {Tian}}, \ and\
  \bibinfo {author} {\bibfnamefont {Y.-Y.}\ \bibnamefont {Liu}},\ }\bibfield
  {title} {\enquote {\bibinfo {title} {Bridges in complex networks},}\ }\href
  {\doibase 10.1103/PhysRevE.97.012307} {\bibfield  {journal} {\bibinfo
  {journal} {Phys. Rev. E}\ }\textbf {\bibinfo {volume} {97}},\ \bibinfo
  {pages} {012307} (\bibinfo {year} {2018})}\BibitemShut {NoStop}%
\bibitem [{\citenamefont {Erd{\H o}s}\ and\ \citenamefont
  {R{\'e}nyi}(1959)}]{ERgraph}%
  \BibitemOpen
  \bibfield  {author} {\bibinfo {author} {\bibfnamefont {P.}~\bibnamefont
  {Erd{\H o}s}}\ and\ \bibinfo {author} {\bibfnamefont {A.}~\bibnamefont
  {R{\'e}nyi}},\ }\bibfield  {title} {\enquote {\bibinfo {title} {On random
  graphs i.}}\ }\href@noop {} {\bibfield  {journal} {\bibinfo  {journal} {Publ.
  Math.}\ }\textbf {\bibinfo {volume} {6}},\ \bibinfo {pages} {290} (\bibinfo
  {year} {1959})}\BibitemShut {NoStop}%
\bibitem [{\citenamefont {Zachary}(1977)}]{Zachary:1977fs}%
  \BibitemOpen
  \bibfield  {author} {\bibinfo {author} {\bibfnamefont {W.~W.}\ \bibnamefont
  {Zachary}},\ }\bibfield  {title} {\enquote {\bibinfo {title} {{An information
  flow model for conflict and fission in small groups}},}\ }\href@noop {}
  {\bibfield  {journal} {\bibinfo  {journal} {J. Anthropol. Res.}\ }\textbf
  {\bibinfo {volume} {33}},\ \bibinfo {pages} {452} (\bibinfo {year}
  {1977})}\BibitemShut {NoStop}%
\bibitem [{\citenamefont {Gabasova}(2016)}]{StarWarssocialnet:2016iy}%
  \BibitemOpen
  \bibfield  {author} {\bibinfo {author} {\bibfnamefont {E.}~\bibnamefont
  {Gabasova}},\ }\href {\doibase 10.5281/zenodo.1411479} {\enquote {\bibinfo
  {title} {{Star Wars social network}},}\ }\bibinfo {howpublished}
  {\url{https://doi.org/10.5281/zenodo.1411479}} (\bibinfo {year}
  {2016})\BibitemShut {NoStop}%
\bibitem [{\citenamefont {G{\'e}nois}\ \emph {et~al.}(2015)\citenamefont
  {G{\'e}nois}, \citenamefont {Vestergaard}, \citenamefont {Fournet},
  \citenamefont {Panisson}, \citenamefont {Bonmarin},\ and\ \citenamefont
  {Barrat}}]{Genois:2015gm}%
  \BibitemOpen
  \bibfield  {author} {\bibinfo {author} {\bibfnamefont {M.}~\bibnamefont
  {G{\'e}nois}}, \bibinfo {author} {\bibfnamefont {C.~L.}\ \bibnamefont
  {Vestergaard}}, \bibinfo {author} {\bibfnamefont {J.}~\bibnamefont
  {Fournet}}, \bibinfo {author} {\bibfnamefont {A.}~\bibnamefont {Panisson}},
  \bibinfo {author} {\bibfnamefont {I.}~\bibnamefont {Bonmarin}}, \ and\
  \bibinfo {author} {\bibfnamefont {A.}~\bibnamefont {Barrat}},\ }\bibfield
  {title} {\enquote {\bibinfo {title} {{Data on face-to-face contacts in an
  office building suggest a low-cost vaccination strategy based on community
  linkers}},}\ }\href@noop {} {\bibfield  {journal} {\bibinfo  {journal} {Netw.
  Sci.}\ }\textbf {\bibinfo {volume} {3}},\ \bibinfo {pages} {326} (\bibinfo
  {year} {2015})}\BibitemShut {NoStop}%
\bibitem [{\citenamefont {Mastrandrea}\ \emph {et~al.}(2015)\citenamefont
  {Mastrandrea}, \citenamefont {Fournet},\ and\ \citenamefont
  {Barrat}}]{Mastrandrea:2015he}%
  \BibitemOpen
  \bibfield  {author} {\bibinfo {author} {\bibfnamefont {R.}~\bibnamefont
  {Mastrandrea}}, \bibinfo {author} {\bibfnamefont {J.}~\bibnamefont
  {Fournet}}, \ and\ \bibinfo {author} {\bibfnamefont {A.}~\bibnamefont
  {Barrat}},\ }\bibfield  {title} {\enquote {\bibinfo {title} {{Contact
  patterns in a high school: A comparison between data collected using wearable
  sensors, contact diaries and friendship surveys}},}\ }\href@noop {}
  {\bibfield  {journal} {\bibinfo  {journal} {PLOS ONE}\ }\textbf {\bibinfo
  {volume} {10}},\ \bibinfo {pages} {e0136497} (\bibinfo {year}
  {2015})}\BibitemShut {NoStop}%
\bibitem [{\citenamefont {Kim}\ \emph {et~al.}(2018)\citenamefont {Kim},
  \citenamefont {Olave-Rojas}, \citenamefont {{\'A}lvarez-Miranda},\ and\
  \citenamefont {Son}}]{Kim:2018ie}%
  \BibitemOpen
  \bibfield  {author} {\bibinfo {author} {\bibfnamefont {H.}~\bibnamefont
  {Kim}}, \bibinfo {author} {\bibfnamefont {D.}~\bibnamefont {Olave-Rojas}},
  \bibinfo {author} {\bibfnamefont {E.}~\bibnamefont {{\'A}lvarez-Miranda}}, \
  and\ \bibinfo {author} {\bibfnamefont {S.-W.}\ \bibnamefont {Son}},\
  }\bibfield  {title} {\enquote {\bibinfo {title} {{In-depth data on the
  network structure and hourly activity of the Central Chilean power grid}},}\
  }\href@noop {} {\bibfield  {journal} {\bibinfo  {journal} {Sci. Data}\
  }\textbf {\bibinfo {volume} {5}},\ \bibinfo {pages} {180209} (\bibinfo {year}
  {2018})}\BibitemShut {NoStop}%
\bibitem [{\citenamefont {Holme}(2018)}]{petterzachary}%
  \BibitemOpen
  \bibfield  {author} {\bibinfo {author} {\bibfnamefont {P.}~\bibnamefont
  {Holme}},\ }\href
  {https://petterhol.me/2018/01/28/zacharys-zachary-karate-club/} {\enquote
  {\bibinfo {title} {{Zachary's Zachary karate club}},}\ }\bibinfo
  {howpublished}
  {\url{https://petterhol.me/2018/01/28/zacharys-zachary-karate-club/}}
  (\bibinfo {year} {2018})\BibitemShut {NoStop}%
\bibitem [{\citenamefont {Jones-Correa}(2012)}]{jones-correa_2012}%
  \BibitemOpen
  \bibfield  {author} {\bibinfo {author} {\bibfnamefont {Michael}\ \bibnamefont
  {Jones-Correa}},\ }\bibfield  {title} {\enquote {\bibinfo {title} {How
  immigrants are marked as outsiders},}\ }\href
  {https://www.nytimes.com/roomfordebate/2012/11/15/how-immigrants-come-to-be-seen-as-americans/how-immigrants-are-marked-as-outsiders}
  {\bibfield  {journal} {\bibinfo  {journal} {The New York Times}\ } (\bibinfo
  {year} {2012})}\BibitemShut {NoStop}%
\bibitem [{\citenamefont {Cowie}\ \emph {et~al.}(2002)\citenamefont {Cowie},
  \citenamefont {Naylor}, \citenamefont {Rivers}, \citenamefont {Smith},\ and\
  \citenamefont {Pereira}}]{Cowie:2002bc}%
  \BibitemOpen
  \bibfield  {author} {\bibinfo {author} {\bibfnamefont {H.}~\bibnamefont
  {Cowie}}, \bibinfo {author} {\bibfnamefont {P.}~\bibnamefont {Naylor}},
  \bibinfo {author} {\bibfnamefont {I.}~\bibnamefont {Rivers}}, \bibinfo
  {author} {\bibfnamefont {P.~K.}\ \bibnamefont {Smith}}, \ and\ \bibinfo
  {author} {\bibfnamefont {B.}~\bibnamefont {Pereira}},\ }\bibfield  {title}
  {\enquote {\bibinfo {title} {{Measuring workplace bullying}},}\ }\href@noop
  {} {\bibfield  {journal} {\bibinfo  {journal} {Aggress. Viol. Behav.}\
  }\textbf {\bibinfo {volume} {7}},\ \bibinfo {pages} {33} (\bibinfo {year}
  {2002})}\BibitemShut {NoStop}%
\bibitem [{\citenamefont {Alivernini}\ \emph {et~al.}(2017)\citenamefont
  {Alivernini}, \citenamefont {Manganelli}, \citenamefont {Cavicchiolo},\ and\
  \citenamefont {Lucidi}}]{Alivernini:2017bl}%
  \BibitemOpen
  \bibfield  {author} {\bibinfo {author} {\bibfnamefont {F.}~\bibnamefont
  {Alivernini}}, \bibinfo {author} {\bibfnamefont {S.}~\bibnamefont
  {Manganelli}}, \bibinfo {author} {\bibfnamefont {E.}~\bibnamefont
  {Cavicchiolo}}, \ and\ \bibinfo {author} {\bibfnamefont {F.}~\bibnamefont
  {Lucidi}},\ }\bibfield  {title} {\enquote {\bibinfo {title} {{Measuring
  bullying and victimization among immigrant and native primary school
  students: Evidence from Italy}},}\ }\href@noop {} {\bibfield  {journal}
  {\bibinfo  {journal} {J. of Psychoeduc. Assess.}\ }\textbf {\bibinfo {volume}
  {37}},\ \bibinfo {pages} {226} (\bibinfo {year} {2017})}\BibitemShut
  {NoStop}%
\bibitem [{\citenamefont {Vivolo-Kantor}\ \emph {et~al.}(2014)\citenamefont
  {Vivolo-Kantor}, \citenamefont {Martell}, \citenamefont {Holland},\ and\
  \citenamefont {Westby}}]{VivoloKantor:2014cca}%
  \BibitemOpen
  \bibfield  {author} {\bibinfo {author} {\bibfnamefont {A.~M.}\ \bibnamefont
  {Vivolo-Kantor}}, \bibinfo {author} {\bibfnamefont {B.~N.}\ \bibnamefont
  {Martell}}, \bibinfo {author} {\bibfnamefont {K.~M.}\ \bibnamefont
  {Holland}}, \ and\ \bibinfo {author} {\bibfnamefont {R.}~\bibnamefont
  {Westby}},\ }\bibfield  {title} {\enquote {\bibinfo {title} {{A systematic
  review and content analysis of bullying and cyber-bullying measurement
  strategies}},}\ }\href@noop {} {\bibfield  {journal} {\bibinfo  {journal}
  {Aggression and Violent Behavior}\ }\textbf {\bibinfo {volume} {19}},\
  \bibinfo {pages} {423} (\bibinfo {year} {2014})}\BibitemShut {NoStop}%
\bibitem [{\citenamefont {Brandwatch}(2019)}]{Peerindex}%
  \BibitemOpen
  \bibfield  {author} {\bibinfo {author} {\bibnamefont {Brandwatch}},\ }\href
  {https://www.brandwatch.com/p/peerindex-brandwatch} {\enquote {\bibinfo
  {title} {{Peer Index}},}\ }\bibinfo {howpublished}
  {\url{https://www.brandwatch.com/p/peerindex-brandwatch}} (\bibinfo {year}
  {2019})\BibitemShut {NoStop}%
\bibitem [{\citenamefont {{Sprout Social}}(2019)}]{SproutSocial}%
  \BibitemOpen
  \bibfield  {author} {\bibinfo {author} {\bibnamefont {{Sprout Social}}},\
  }\href {https://sproutsocial.com} {\enquote {\bibinfo {title} {{Sprout
  Social}},}\ }\bibinfo {howpublished} {\url{https://sproutsocial.com}}
  (\bibinfo {year} {2019})\BibitemShut {NoStop}%
\bibitem [{\citenamefont {Holme}\ and\ \citenamefont
  {Saram{\"a}ki}(2012)}]{TemporalNetwork}%
  \BibitemOpen
  \bibfield  {author} {\bibinfo {author} {\bibfnamefont {P.}~\bibnamefont
  {Holme}}\ and\ \bibinfo {author} {\bibfnamefont {J.}~\bibnamefont
  {Saram{\"a}ki}},\ }\bibfield  {title} {\enquote {\bibinfo {title} {Temporal
  networks},}\ }\href {\doibase https://doi.org/10.1016/j.physrep.2012.03.001}
  {\bibfield  {journal} {\bibinfo  {journal} {Physics Reports}\ }\textbf
  {\bibinfo {volume} {519}},\ \bibinfo {pages} {97} (\bibinfo {year} {2012})},\
  \bibinfo {note} {temporal Networks}\BibitemShut {NoStop}%
\bibitem [{\citenamefont {Ahn}\ \emph {et~al.}(2010)\citenamefont {Ahn},
  \citenamefont {Bagrow},\ and\ \citenamefont {Lehmann}}]{Ahn:2010dj}%
  \BibitemOpen
  \bibfield  {author} {\bibinfo {author} {\bibfnamefont {Y.-Y.}\ \bibnamefont
  {Ahn}}, \bibinfo {author} {\bibfnamefont {J.~P.}\ \bibnamefont {Bagrow}}, \
  and\ \bibinfo {author} {\bibfnamefont {S.}~\bibnamefont {Lehmann}},\
  }\bibfield  {title} {\enquote {\bibinfo {title} {{Link communities reveal
  multiscale complexity in networks}},}\ }\href@noop {} {\bibfield  {journal}
  {\bibinfo  {journal} {Nature}\ }\textbf {\bibinfo {volume} {466}},\ \bibinfo
  {pages} {761} (\bibinfo {year} {2010})}\BibitemShut {NoStop}%
\bibitem [{\citenamefont {Jeub}\ \emph {et~al.}(2018)\citenamefont {Jeub},
  \citenamefont {Sporns},\ and\ \citenamefont {Fortunato}}]{Jeub:2018ik}%
  \BibitemOpen
  \bibfield  {author} {\bibinfo {author} {\bibfnamefont {L.~G.~S.}\
  \bibnamefont {Jeub}}, \bibinfo {author} {\bibfnamefont {Olaf}\ \bibnamefont
  {Sporns}}, \ and\ \bibinfo {author} {\bibfnamefont {Santo}\ \bibnamefont
  {Fortunato}},\ }\bibfield  {title} {\enquote {\bibinfo {title}
  {{Multiresolution consensus clustering in networks}},}\ }\href@noop {}
  {\bibfield  {journal} {\bibinfo  {journal} {Sci. Rep.}\ }\textbf {\bibinfo
  {volume} {8}},\ \bibinfo {pages} {3259} (\bibinfo {year} {2018})}\BibitemShut
  {NoStop}%
\bibitem [{\citenamefont {Gates}\ and\ \citenamefont
  {Ahn}(2017)}]{Gates:2017ta}%
  \BibitemOpen
  \bibfield  {author} {\bibinfo {author} {\bibfnamefont {A.~J.}\ \bibnamefont
  {Gates}}\ and\ \bibinfo {author} {\bibfnamefont {Y.-Y.}\ \bibnamefont
  {Ahn}},\ }\bibfield  {title} {\enquote {\bibinfo {title} {{The impact of
  random models on clustering similarity}},}\ }\href@noop {} {\bibfield
  {journal} {\bibinfo  {journal} {J. Mach. Learn. Res.}\ }\textbf {\bibinfo
  {volume} {18}},\ \bibinfo {pages} {1} (\bibinfo {year} {2017})}\BibitemShut
  {NoStop}%
\end{thebibliography}%

\end{document}